\documentclass[journal]{IEEEtran} 
\IEEEoverridecommandlockouts
\usepackage{cite}
\usepackage{comment}
\usepackage{amsmath,amssymb,amsfonts}
\usepackage{algorithmic}
\usepackage{graphicx}
\usepackage{textcomp}
\usepackage{xcolor}
\usepackage{tabularx}
\usepackage{adjustbox}
\usepackage{tabularx}
\usepackage{booktabs}
\usepackage{tablefootnote}
\def\BibTeX{{\rm B\kern-.05em{\sc i\kern-.025em b}\kern-.08em
    T\kern-.1667em\lower.7ex\hbox{E}\kern-.125emX}}

\usepackage{subfigure}
    
\begin{document}

\title{Humans are Still Better than ChatGPT: \\Case of the IEEEXtreme Competition
\thanks{}
}



\author{
\IEEEauthorblockN{Anis Koubaa\IEEEauthorrefmark{1}, Basit Qureshi, Adel Ammar, Zahid Khan, Wadii Boulila, Lahouari Ghouti\\}
\IEEEauthorblockA{\textit{Prince Sultan University, Saudi Arabia\\}
}
\IEEEauthorblockA{\IEEEauthorrefmark{1}akoubaa@psu.edu.sa}
}

\maketitle

\begin{abstract}
Since the release of ChatGPT, numerous studies have highlighted the remarkable performance of ChatGPT, which often rivals or even surpasses human capabilities in various tasks and domains. However, this paper presents a contrasting perspective by demonstrating an instance where human performance excels in typical tasks suited for ChatGPT, specifically in the domain of computer programming. We utilize the IEEExtreme Challenge competition as a benchmark—a prestigious, annual international programming contest encompassing a wide range of problems with different complexities. To conduct a thorough evaluation, we selected and executed a diverse set of 102 challenges, drawn from five distinct IEEExtreme editions, using three major programming languages: Python, Java, and C++. Our empirical analysis provides evidence that contrary to popular belief, human programmers maintain a competitive edge over ChatGPT in certain aspects of problem-solving within the programming context. In fact, we found that the average score obtained by ChatGPT on the set of IEEExtreme programming problems is 3.9 to 5.8 times lower than the average human score, depending on the programming language. This paper elaborates on these findings, offering critical insights into the limitations and potential areas of improvement for AI-based language models like ChatGPT.

\end{abstract}


\section{Introduction}
\subsection{Background and motivation}

Large Language Models (LLMs)\cite{zhao2023survey} have emerged as a groundbreaking artificial intelligence technology, especially since the release of ChatGPT in late November 2022. LLMs can mimic human-level capabilities in various complex natural language processing and understanding tasks across multiple domains, such as virtual assistants, chatbots, language translation, sentiment analysis, and more. ChatGPT has been trained on an extensive corpus of data spanning various disciplines, enabling it to acquire a broad spectrum of knowledge. Its training data comprises diverse sources from multiple domains, including but not limited to science, literature, law, programming, finance, and many more. This various training data has given ChatGPT a global perspective, making it capable of understanding and generating responses across a wide range of subjects. The vast knowledge base of ChatGPT allows it to provide insights and solutions to complex problems that span different domains, making it an effective tool for various applications in natural language processing and understanding.

With the ChatGPT's unprecedented capabilities compared to other LLMs in competing with humans across various applications, there has been a significant increase in the number of studies investigating its performance in specialized and complex domains, such as healthcare \cite{qiu2023large}, and finance \cite{wu2023bloomberggpt}. However, despite the growing interest in evaluating ChatGPT's abilities in these areas, there has been a lack of research specifically focusing on its specific performance in problem-solving and programming assessment domains, which is the main focus of this paper. This research gap has motivated us to investigate and evaluate ChatGPT's abilities in these areas.

\subsection{Objective}
This paper aims to investigate the problem-solving capabilities of ChatGPT by evaluating its performance on programming problem benchmarks. Our objective is to assess how ChatGPT compares to human programmers and to extract valuable insights into its strengths and weaknesses in this domain-specific context.  

To accomplish our research objective, we identified the IEEExtreme Programming Challenge as the most reputable and prestigious competition that could serve as an appropriate benchmark for comparing the problem-solving abilities of ChatGPT and human programmers. The IEEExtreme Programming Challenge is an annual international programming competition organized by the Institute of Electrical and Electronics Engineers (IEEE). This 24-hour competition attracts programming professionals from across the globe to compete in solving programming problems with varying degrees of complexity, which demand high-level problem-solving and programming skills.

In summary, the primary objective of this study is to evaluate the performance of ChatGPT in the programming and problem-solving-specific context. To this end, we will employ the IEEExtreme Challenge competition as a benchmark, utilizing problems of varying complexities. Moreover, we aim to analyze the limitations of ChatGPT in solving specific problems and programming tasks and identify areas for improvement and optimization. By conducting this analysis, we aim to provide the community with insights into the effectiveness of ChatGPT in programming and problem-solving domains and provide recommendations for future developments in this area. 

\subsection{Methodology}
For this study, we selected five IEEExtreme programming competitions, each consisting of an average of 20 questions. To guide ChatGPT in designing solutions while meeting non-functional requirements such as memory usage and execution time, we designed well-crafted prompts. For each problem, we presented the prompt to ChatGPT and evaluated its corresponding solution using Hackerrank, which was used to generate scores. We evaluated solutions in three top programming languages: Python 3, C++ 11, and Java 7. In case of errors, we made up to seven attempts to guide ChatGPT toward the correct solution by providing the corresponding Hackerrank error message. 

Furthermore, to ensure the consistency of the results, we conducted this process three times, using different ChatGPT chat windows for each programming problem of the five selected IEEExtreme Challenges. The final results were analyzed, and we identified ChatGPT's limitations in solving specific problems and programming tasks. Additionally, we provided recommendations for areas of improvement and optimization, which could enhance ChatGPT's effectiveness in programming and problem-solving domains.

\subsection*{Research Questions}
In this study, we aim to respond to four research questions:
\begin{enumerate}
\item How does ChatGPT compare to human programmers in problem-solving and programming tasks, in the context of the IEEExtreme Challenge competition?
\item In which specific programming tasks or problem types do humans outperform ChatGPT, and what are the underlying reasons for this disparity?
\item Is ChatGPT performance biased towards particular programming languages among the three selected languages, namely, Python, C++ 11, and Java?
\item What are the fundamental limitations of ChatGPT in programming and problem-solving, and how can these findings guide future research and development of ChatGPT and other domain-specific large language models?
\end{enumerate}

\subsection{Overview of the paper structure}

The paper is organized as follows. Section II provides a literature review on ChatGPT and its applications, human performance in programming tasks, and previous comparisons between AI and human performance. Section III describes the methodology, including selecting IEEExtreme challenges, evaluation criteria and metrics, programming languages used, and the data collection and analysis approach. Section IV presents the results of the study that compares ChatGPT and human performance in programming tasks of IEEExtreme challenges and identifies the gap with human-level performance. The interpretation of the results and limitations of ChatGPT in programming and problem-solving tasks, as well as the reasons for disparities, are discussed in the same section. Finally, Section V concludes the paper with a summary of findings, potential areas for improvement in ChatGPT, and suggestions for future research directions.

\section{Related Works}

\subsection{ChatGPT and its applications}
ChatGPT has made significant progress and it has been used in various applications. The detailed comparison of ChatGPT performance in various domains is shown in Table \ref{tab:related_work}. In our previous study, ChatGPT's applications were classified into five main categories \cite{koubaa2023exploring}:
\begin{itemize}
    \item \textbf{NLP:} In this type of application, ChatGPT generates human-like responses in natural language. Applications of ChatGPT in NLP include building virtual assistants, chatbots, language translation systems, and text generation tasks such as summarization and question answering \cite{guo2023close,bang2023multitask,ortega2023linguistic}.
    \item \textbf{Healthcare: }ChatGPT has been used in various healthcare fields. It has been applied in healthcare decision support to provide relevant information and recommendations \cite{liu2023assessing}. In addition, many recent research works have investigated the case of using ChatGPT in patient education, where ChatGPT provides patients with educational information about their health conditions, treatments, and medications \cite{sallam2023chatgpt}. Moreover, ChatGPT has been included in applications related to telemedicine to provide more efficient and accurate virtual diagnosis and treatment \cite{iftikhar2023docgpt}.
    \item \textbf{Ethics:} Many recent research works addressed the challenge of using ChatGPT for the benefit of society and how to maintain public safety \cite{hacker2023regulating}. Many authors explored using ChatGPT to generate student works and scientific publications \cite{fergus2023evaluating}. Many other researchers focused on the ethical concerns, data biases, and safety issues related to ChatGPT \cite{cotton2023chatting}.
    \item \textbf{Education:} ChatGPT has played an essential role in several applications in education \cite{frieder2023mathematical,tlili2023if,Qureshi2023}. It helped to improve the learning experience for students. It can provide personalized educational content. In addition, it can generate educational materials for students and tutors. ChatGPT is considered a promising tool for education, as it can provide insightful direction and feedback.
    \item \textbf{Industry:} recently, various applications across many industries have been focused on using ChatGPT to improve efficiency, streamline processes, and enhance customer experiences \cite{prieto2023investigating,dowling2023chatgpt,kocaballi2023conversational}. Applications include the manufacturing industry, where it can monitor and control production processes. In addition, ChatGPT is used in the financial sector, where it can offer support to customers and company owners. Moreover, it can provide customer support to handle routine inquiries.
    \end{itemize}


\subsection{Previous comparisons between AI and human performance}

In a recent study \cite{guo2023close}, Guo et al. compared the responses of ChatGPT and human experts to around 40K questions in various domains, such as finance, psychology, medical, legal, and open-domain, in both English and Chinese languages. They analyzed ChatGPT's response characteristics, differences and gaps from human experts, and future directions for LLMs. The researchers discovered that ChatGPT's responses are generally more helpful than human experts' in over half of the questions, especially in finance and psychology, due to its ability to offer specific suggestions. However, ChatGPT performs poorly in the medical domain. The authors also found that ChatGPT writes in an organized manner, with clear logic, and tends to provide detailed answers with less bias and harmful information. However, it may fabricate facts. Notably, the study did not include programming tasks but only theoretical questions about computer science-related concepts taken from Wikipedia.

On another hand, Qin et al. \cite{qin2023chatgpt} examined the zero-shot learning ability of ChatGPT. The evaluation was conducted on 20 commonly used natural language processing (NLP) datasets covering seven task categories, including natural language inference, question answering, dialogue, summarization, named entity recognition, and sentiment analysis. However, the study did not include any programming tasks. The authors performed extensive empirical studies to analyze the strengths and limitations of the current version of ChatGPT. They discovered that ChatGPT performed well on tasks that require reasoning abilities, such as arithmetic reasoning, but it struggled with specific tasks like sequence tagging. Additionally, ChatGPT was outperformed by previous models that had been fine-tuned for a specific task. The findings suggest that ChatGPT is still far from reaching perfection as a generalist model.

Kashefi and Mukerji \cite{kashefi2023chatgpt} investigated the potential of ChatGPT in one specific aspect of programming, which is to produce numerical algorithms for solving mathematical problems. They explored generating code in different programming languages, debugging user-written code, completing unfinished code, rewriting code in different programming languages, and parallelizing serial code. Although the study outcomes demonstrated that ChatGPT is capable of programming numerical algorithms, certain limitations and challenges were encountered. These included issues such as generating singular matrices, producing incompatible arrays, and irregular interruption when generating long codes required for scientific simulations. Another challenge was the inclusion of unknown libraries. Despite these limitations, the study suggests that ChatGPT has the potential for further development and improvement in programming numerical algorithms.

Liu et al. \cite{liu2023evaluating} evaluated the performance of ChatGPT and GPT-4 on various logical reasoning tasks using multiple datasets, including both well-known benchmarks and newly-released ones. The experiments showed that ChatGPT performs better than the RoBERTa \cite{liu2019roberta} fine-tuning method on most logical reasoning benchmarks. However, both ChatGPT and GPT-4 struggle with newly-released and out-of-distribution datasets. GPT-4 showed higher performance than ChatGPT on most logical reasoning datasets. Nevertheless, despite advancements in models like ChatGPT and GPT-4, the task of logical reasoning still poses significant challenges for these models, particularly when dealing with out-of-distribution and natural language inference datasets.

Tian et al. \cite{tian2023chatgpt} presented an empirical study evaluating the potential of the ChatGPT generative large-scale language model as an assistant bot for programmers. The study assesses ChatGPT's performance on three code-related tasks: code generation, program repair, and code summarization. ChatGPT is found to perform well in the code generation task but struggles to generalize to new and unseen problems. The study also highlights the negative impact of long prompts on ChatGPT's inference capabilities. In the program repair task, ChatGPT achieves competitive results compared to Refactory \cite{hu2019re}, a state-of-the-art semantic-based assignments repair tool. However, prompts that are not related to bug information are found to make ChatGPT perform even worse due to its limited attention span. The study's limitation pertains to the absence of a comparison between ChatGPT's performance and that of human programmers, thus hindering the establishment of its proficiency in relation to human experts.

Biswas \cite{biswas2023role} explored existing language models and tools for computer programming. ChatGPT is introduced as a powerful and versatile tool that can perform a variety of programming-related tasks such as code completion, correction, optimization, and refactoring. The paper highlights the ability of ChatGPT to provide explanations and guidance to help users understand complex concepts and resolve technical issues. The use of ChatGPT is noted as a potential means to improve overall satisfaction with support services and build a reputation for expertise and reliability. In summary, the paper suggests that ChatGPT is a valuable resource for technical support and improving efficiency and accuracy in computer programming tasks. The author solved simple programs without comparing them with human performance. Additionally, the author's work is only limited to exploratory analysis without empirical results. 

In reference \cite{avila2023chatgpt}, the authors discussed the challenges faced by behavior analysts in automating and systematizing experimental tasks. With the development of online platforms, OpenAI ChatGPT has emerged as a chatbot that can generate text responses similar to humans in a conversational context. One of its key functions is the ability to generate programming code blocks in various programming languages. The article presents the use of ChatGPT as a programming assistant to develop an online behavioral task using HTML, CSS, and JavaScript code. While ChatGPT cannot replace programmers entirely, it can provide detailed programming solutions and reduce the time associated with programming. The authors assess the performance of the ChatGPT with random problems in diverse directions. There is no quantitative study to assess the performance of the ChatGPT. It also lacks comparison with human performance. 

\begin{table*}[ht]
    \centering
    \caption{Related work on ChatGPT}
    \begin{tabular}{|c|c|p{5cm}|p{5cm}|}
        \hline
        \textbf{Paper} & \textbf{Approach} & \textbf{Main Finding} & \textbf{Limitations} \\ 
        \hline
        Guo et al. \cite{guo2023close} & NLP & ChatGPT is more helpful than human experts in finance and psychology questions. & Poor performance in the medical domain; may fabricate facts. \\
        \hline
        Qin et al. \cite{qin2023chatgpt} & NLP & ChatGPT performs well on reasoning tasks but struggles with specific tasks. & Outperformed by models fine-tuned for specific tasks. \\
        \hline
        Kashefi and Mukerji \cite{kashefi2023chatgpt} & Programming & ChatGPT can program numerical algorithms but faces challenges generating long codes and using unknown libraries. & Issues with singular matrices and incompatible arrays. \\
        \hline
        Liu et al. \cite{liu2023evaluating} & Logical Reasoning & ChatGPT outperforms RoBERTa on most benchmarks but struggles with newly-released datasets. & Challenge in dealing with out-of-distribution and natural
language inference datasets \\
\hline 
Tian et al. \cite{tian2023chatgpt} & Programming & highlights the ability of ChatGPT to provide explanations and guidance to help users understand complex concepts and resolve technical issues. & No comparison with human programmers to evaluate ChatGPT's performance relative to human experts. \\
\hline
Biwas et al. \cite{biswas2023role} & Programming & ChatGPT performs well in the code generation task but struggles to generalize to new and unseen problems. & The study is only limited to exploratory analysis without empirical results. No comparison with human performance \\
\hline
Avila et al. \cite{avila2023chatgpt} & Programming & The use of ChatGPT as a programming assistant to develop an online behavioral task using HTML, CSS, and JavaScript code. & There is no quantitative study to assess the performance of ChatGPT. It also lacks comparison with human performance. \\
        \hline
    \end{tabular}
    \label{tab:related_work}
\end{table*}

The previous studies on ChatGPT mainly explored its performance in various contexts, but most of them did not follow a quantitative approach. In contrast, our study quantitatively evaluates ChatGPT's performance in solving IEEE Xtreme problems and compares it to average human performance in three different programming languages.


\section{Methodology}

\subsection{The IEEExtreme Competition}
There are several global programming competitions including, IEEE Extreme Programming Competition (IEEEXtreme) \cite{IEEExtreme}, ACM International Collegiate Programming Contest (ICPC) \cite{ACMICPC}, Google Code Jam \cite{GoogleCodeJam}, Facebook / Meta Hacker Cup \cite{MetaHackerCup}, and International Olympiad in Informatics (IOI) \cite{IOI}, to name a few. The IEEEXtreme is a global programming competition organized by the Institute of Electrical and Electronics Engineers (IEEE). The IEEEXtreme programming competition has been running annually since 2006. The number of participants in the competition has been increasing each year. In the early years, the competition had around 500 teams participating. In recent years, the number of participating teams has grown to around 10,000 or more, with participants from over 100 countries. The competition has become a major event in the global programming community, attracting top talent worldwide. The competition provides a platform for students to showcase their technical skills and talent. Winning the competition is a significant achievement that can help participants stand out to potential employers or graduate schools.

It is a 24-hour coding marathon where teams of up to three students worldwide compete to solve a series of challenging programming problems. The problems are usually related to topics in computer science, mathematics, and engineering. The competition is designed to encourage and develop programming skills in students, as well as to promote teamwork and creativity. Participants must rely on their knowledge of algorithms, data structures, and programming languages to solve problems within the time limit. The competition is judged based on the number of problems solved, with ties broken based on the time taken to solve them. Each problem has a set number of test cases defined in the evaluation platform. Students need to provide a programming solution to the given problem by passing all the test cases to earn scores. 

\subsection{Selecting IEEExtreme Challenges}
IEEExtreme competition editions 11 and beyond were held on the CSacademy platform \cite{CSacademy}, while the earlier editions were conducted on the Hackerrank platform \cite{Hackerrank}. These platforms are accessible worldwide and open to all, and both may rely on Amazon Web Services (AWS) for their infrastructure. The entire problem set for IEEExtreme competitions versions 15 and 16, which were hosted in 2021 and 2022 respectively, are available on the CSacademy platform. The earlier versions are unfortunately not available. Selected problems from versions 8, 9, and 10 are available on the Hackerrank platform, however, these can be accessed using the practice community website only. In this work, we rely on the problem sets presented in IEEExtreme versions 8, 9, 10, 15, and 16. Table \ref{tab:scores_per_language} shows a list of problems presented in each of these competitions. Each problem is classified based on difficulty level defined by the organizers as easy, medium, hard, and advanced. While all problems are available for IEEExtreme versions 9, 15, and 16, only a select few are available for versions 8 and 10, on the hackerrank \cite{HackerrankCommunity} practice community website.

\begin{table}[h]
\centering
\caption{Classification of challenges in IEEExtreme Competitions}
\label{tab:scores_per_language}
\begin{tabular}{p{14mm} cccccc}

\hline
\textbf{Competition Version} & \textbf{Total} & \textbf{Available} & \textbf{Easy}& \textbf{Medium} &\ \textbf{Hard} & \textbf{Advanced}\\
\hline
Extreme 8 & 23 & 4 & 1 & 3 & 0 & 0\\ 
Extreme 9 & 29 & 29 & 5 & 8 & 8 & 8\\ 
Extreme 10 & 24 & 17 & 0 & 0 & 17 & 0\\ 
Extreme 15 & 26 & 26 & 8 & 10 & 8 & 0\\ 
Extreme 16 & 26 & 26 & 9 & 10 & 7  & 0\\
\hline
\end{tabular}
\end{table}

\subsection{Solving and scoring a problem}
As mentioned in the previous section, each competition comprises a certain number of problems. A problem statement generally includes a brief description of the problem, its input and output format, and any constraints or limitations that apply to the solution. The problem may also include sample test cases that provide examples of the expected input and output. Participants are expected to write a computer program using a programming language of their choice, that solves the problem statement and produces the expected output. The solution is then submitted to the competition's online system, which tests the program against a variety of test cases and assigns a score based on its accuracy and efficiency. In recent versions of the competition, participants are expected to submit correct solutions that satisfy the minimum execution time and memory limitations. This usually requires participants to post optimized solutions based on the correct choice of programming constructs, efficient data structures, and algorithms. 

To obtain scores for each problem, participants can submit solutions multiple times to pass most, if not all, of the hidden test cases. However, multiple submissions to the same problem typically result in point deductions, known as penalties, which are factored into the total score in case of tiebreakers between teams scoring the exact same number of points. A Programmer may solve any of these problems using a programming language of their choice including but not limited to C/C++, java, python, etc. At the end of the competition, the website displays relevant information for each problem, including the average score earned per team and the percentage of teams that attempted to solve the problem. This important information factors into Human performance in this research work and is compared with AI performance, as explained in the next section.

\subsection{ChatGPT Code Generation and Data Collection}
As the IEEExtreme programming competition is open to participants worldwide, it is essential that each participant has access to the necessary resources to solve the problems. With this goal in mind, we have developed a method for participants in this study to simulate the experience of competing in IEEExtreme by using ChatGPT to solve the problems. We evaluated the problem-solving performance of ChatGPT and human programmers using three primary programming languages: Python 3, Java 7, and C++. These languages were selected based on their popularity in the programming community and their suitability for solving the programming problems provided by the IEEExtreme Challenge competition. We ensured that all participants were familiar with these languages and had equivalent levels of proficiency.

As a problem appears, the participants simply copy and paste the problem statement from the competition website into the prompt space. ChatGPT then generates detailed results that include an explanation of the algorithm, complete executable code in the selected programming language, sample test cases, and other relevant information needed to solve the problem. If for any of the following reasons, the provided code fails to execute, the participant will repeat the process for a maximum number of 7 trials until either the code works perfectly, or the number of trials has exhausted. To this end, several prompts would be used to improve the quality of results generated by chatGPT. The reasons include:
\begin{itemize}
\item Incomplete code produced
\item Compile Errors
\item Runtime Errors
\item Memory Limit Exceeded
\item The execution time limit Exceeded
\item Failing test cases
\end{itemize}

The following prompts are to be used to improve the quality of the response generated:
\begin{itemize}
\item Provide a complete code for this problem using [language].
\item Provide an optimized code using [language] that runs the program in a minimum time of x minutes and memory limitations of y Megabytes.
\item Following up on this code, improve it to solve this test case: [provided test case with the expected output generated for certain input].
\end{itemize}

The participant runs the code generated by ChatGPT on the platform and records the success rate, which includes the number of passed test cases and the maximum score earned for each problem. This procedure is replicated for all problems, utilizing all three programming languages. The results are recorded by all participants in a shared data repository. The data collected include:
\begin{itemize}
\item Competition edition
\item Problem title identifier
\item Difficulty level of the problem
\item Language used (C++, Java or Python)
\item Number of trials/iterations
\item Scored earned by chatGPT generated code
\item Average Human performance for the problem
\end{itemize}

Data is gathered from all five editions of the IEEExtreme competition, encompassing a total of 102 problems. For every problem set, code is generated using each of the three programming languages, with at least three iterations performed. Participants conducted three to seven trials per iteration to complete the execution of the generated code. The data is analyzed to generate conclusions for this study. These are presented in the next sections.

\section{Discussion}
\subsection{Interpretation of the results}

 Table \ref{tab:scores_per_language} shows the average score that ChatGPT achieved on the set of programming problems for Python 3, Java 7, and C++ 11, as well as the average human performance on the same set of problems. It appears that ChatGPT's average score is significantly (from 3.9 to 5.8 times) lower than the average human performance for all three languages, which shows that there is still a large room for improvement in ChatGPT's programming abilities. It is noteworthy that ChatGPT's performance varies among the three programming languages, with Java 7 showing the highest average score, followed by Python 3 and then C++. This observation may suggest that the size and quality of available learning materials for each language in ChatGPT's dataset are not equal. Indeed, Java has been the most widely used language for many years and has extensive documentation, which could have contributed to ChatGPT's better performance on problems written in Java 7 compared to Python 3 and C++.
 
\begin{table}[h]
\centering
\caption{Average ChatGPT and Human Scores on the IEEEXtreme problems per Programming Language}
\label{tab:scores_per_language}
\begin{tabular}{lcc}
\hline
\textbf{Language} & \textbf{Avg ChatGPT Score} & \textbf{Avg Human Performance} \ \\
\hline
Python 3 & 9.06 & 44.5 \\ \hline
Java 7 & 11.46 & 44.5 \\ \hline
C++ & 7.67 & 44.5 \\ 
\hline
\end{tabular}
\end{table}

\begin{figure}[h]
\centering
\includegraphics[width=0.9\columnwidth]{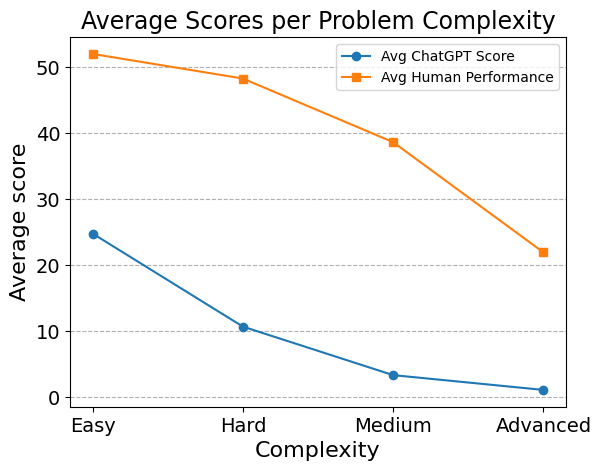}
\caption{Comparison of average ChatGPT and human scores by problem complexity, in the IEEEXtreme competition.}
\label{fig:Scores_per_complexity}
\end{figure}

Figure \ref{fig:Scores_per_complexity} shows the average ChatGPT and human performances for programming problems categorized by their complexity level. The complexity levels are Easy, Hard, Medium, and Advanced. As expected, when the complexity level of the problems increases, both ChatGPT's average score and human performance significantly decrease. However, the decrease is much sharper for ChatGPT. Its score is 23 times lower for the Advanced category compared to the Easy category, while this decrease is only of 2.4 times for human performance. It should be noted that the categories 'Hard' and 'Medium' used by IEEEXtreme competition may not be accurate indicators of problem difficulty, as both humans and ChatGPT demonstrate significantly better performance in the 'Hard' category compared to the 'Medium' category. This also highlights the subjective character of this categorization. On another hand, the correlation coefficient between ChatGPT's and human scores is low (0.21), which indicates that the easiest programming problems for human programmers are not necessarily the easiest for ChatGPT to solve, and vice versa. This lack of correlation between ChatGPT's and human scores could be due to various factors, such as differences in problem-solving approaches and strategies used by ChatGPT and human programmers, variations in the level of programming expertise, and the type and complexity of the problems presented to them.


Table \ref{tab:scores_per_language_and_complexity} breaks down further these results per complexity levels. We notice again that ChatGPT performs better on Java for all complexity levels except the 'Advanced' category. For this category, all the tests on all problems completely failed except one test on the "Finite Domain Constraints" problem from Xtreme9 that gave a partial success (12.74\%) only on Python. Therefore, we cannot draw a general conclusion from this single partial success. On another hand, the average human score presented in this table is the same for all three languages because it was provided by the IEEEXtreme website as an overall average over all programming languages, and no average scores per programming languages was available.

\begin{table}[h]
\centering
\caption{Average ChatGPT and Human Scores on the IEEEXtreme problems per programming language and complexity category}
\label{tab:scores_per_language_and_complexity}
\begin{tabular}{cccc}
\toprule
\textbf{Language} & \textbf{Complexity} & \textbf{\begin{tabular}[c]{@{}c@{}}Avg \\ ChatGPT \\ Score\end{tabular}} & \textbf{\begin{tabular}[c]{@{}c@{}}Avg \\ Human\\ Score\end{tabular}} \\ \hline
\midrule

Python 3 & Easy & 22.17 & 51.96 \\
& Hard & 8.69 & 46.70 \\
& Medium & 1.53 & 38.99 \\
& Advanced & 3.19 & 21.89 \\
\midrule
Java 7
& Easy & 30.06 & 51.96 \\
& Hard & 9.21 & 46.70 \\
& Medium & 4.24 & 38.99 \\
& Advanced & 0.00 & 21.89 \\
\midrule
C++
& Easy & 22.02 & 51.96 \\
& Hard & 4.35 & 46.70 \\
& Medium & 4.64 & 38.99 \\
& Advanced & 0.00 & 21.89 \\
\bottomrule
\end{tabular}
\end{table}

\begin{figure}[h]
\centering
\includegraphics[width=0.8\columnwidth]{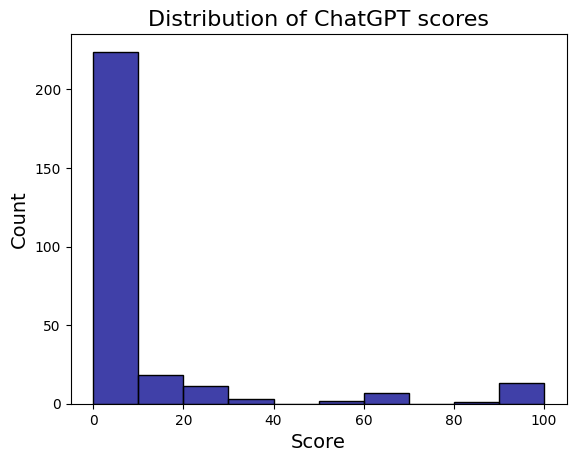}
\includegraphics[width=0.8\columnwidth]{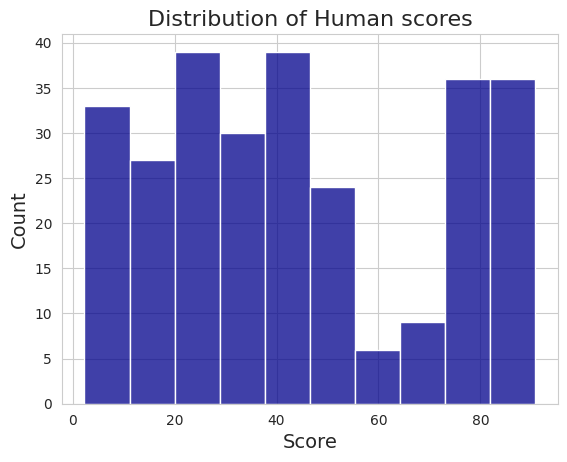}
\caption{Histogram of the Distribution of ChatGPT scores (top) and human programmers' average scores (bottom).}
\label{fig:hist-chatgpt-and-human}
\end{figure}

\begin{figure*}[h]
\includegraphics[width=1.06\textwidth]{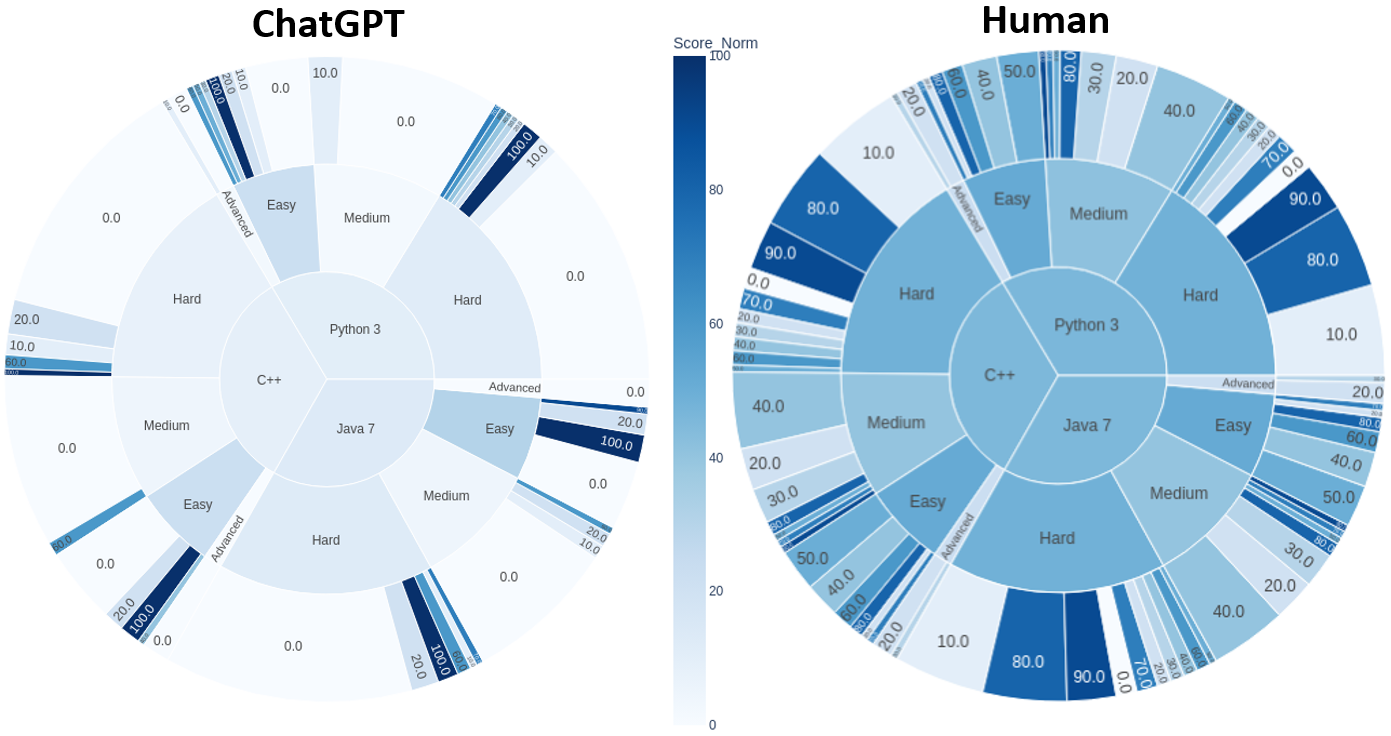}
\caption{Sunburst charts comparing programming language proficiency scores between ChatGPT (left) and human programmers (right) across different complexity levels and programming languages. The scores in the outer sectors have been rounded to the nearest 10.}
\label{fig:sunburst-chatgpt-vs-human}
\end{figure*}

Figure \ref{fig:hist-chatgpt-and-human} shows the complete score distributions of ChatGPT and average human programmers. This figure clearly demonstrates the marked superiority of average human programmers over ChatGPT, with ChatGPT obtaining a null score in the large majority of cases (72\%), while only 10.0\% of cases correspond to an average human performance less than 10\%. Figure \ref{fig:sunburst-chatgpt-vs-human} compares the sunburst charts of ChatGPT and human scores per programming language and complexity level. The color of inner sectors (representing programming languages and complexity categories) corresponds to the average colors of outer sectors belonging to them. The darker the color, the better the results. This figure provides additional evidence of the dominance of average human programmers over ChatGPT in almost all tested cases.


To get an idea about the progress achieved in GPT-4 compared to GPT-3.5, in terms of programming capabilities, we tested their performance on a representative set of 6 problems using the Python 3 language. The results are presented in Table \ref{tab:gpt_3.5_vs_4}. GPT-4 showed a slight improvement in one problem ("Counting Molecules") with an average score increasing from 65\% to 70\% (but still lower than average human score), and a clear improvement in another problem ("Painter's Dilemma") where it went from complete failure to complete success. However, for the remaining 4 problems, both GPT-3.5 and GPT-4 obtained a score of zero. For the "Painter's Dilemma", which is an optimization problem, we also prompted ChatGPT to generate C++ and Java solutions. Both tests yielded a 100\% success in GPT-4, compared to 4.76\% and 0\%, respectively, in GPT-3.5. These results indicate that the improvement in GPT-4 programming abilities, compared to GPT-3.5 is limited to specific types of problems.

\begin{table}[h]
\centering
\caption{Comparison of GPT-3.5 and GPT-4 average scores on a set of selected problems, using Python 3 language.}
\label{tab:gpt_3.5_vs_4}
\begin{tabular}{p{1.4cm}p{1.1cm}p{1.2cm}ccc}
\hline
\textbf{Problem Title} & \textbf{\begin{tabular}[c]{@{}c@{}}Extreme \\ Edition\end{tabular}} & \textbf{Complexity} & \textbf{\begin{tabular}[c]{@{}c@{}}GPT-3.5 \\ Score\end{tabular}} & \textbf{\begin{tabular}[c]{@{}c@{}}GPT-4 \\ Score\end{tabular}} & \textbf{\begin{tabular}[c]{@{}c@{}}Human \\ score\end{tabular}} \\ \hline
Back to Square 1 & Xtreme8 & Medium & 0.0 & 0.0 & 77.00 \\\hline
Prediction Games & Xtreme9 & Medium & 0.0 & 0.0 & 15.38 \\\hline
Counting Molecules & Xtreme10 & Hard & 65.0 & 70.0 & 84.70 \\\hline
Painter's Dilemma & Xtreme10 & Hard & 0.0 & 100.0 & 80.39 \\\hline
Travel Service & Xtreme16 & Medium & 0.0 & 0.0 & 38.89 \\\hline
My Treat & Xtreme16 & Easy & 0.0 & 0.0 & 58.49 \\ 
\hline 
\end{tabular}
\end{table}

\subsection{Limitations of ChatGPT in programming tasks}
Based on the results of our experiments, we can draw several general conclusions about the limitations of ChatGPT in programming tasks. First, ChatGPT's performance on programming tasks is significantly lower than that of an average human programmer, indicating that there is still a way to go before ChatGPT can fully match human intelligence in programming. This is especially true for more complex problems, where the performance gap between ChatGPT and humans is even more significant. This suggests that ChatGPT still has limitations in understanding and solving complex programming problems that require high-level reasoning and expertise. Second, the lack of correlation between ChatGPT's and human scores indicates that the easiest programming problems for human programmers are not necessarily the easiest for ChatGPT to solve, and vice versa. This suggests that ChatGPT may have limitations in problem-solving approaches and strategies that differ from those of human programmers. Finally, although there have been some improvements in the GPT-4 compared to GPT-3.5 in terms of programming capabilities, there is still a significant performance gap between ChatGPT and human programmers, especially for more complex problems. This suggests that there are still limitations in the current state-of-the-art language models for programming tasks, and that further research and development are needed to bridge the gap between ChatGPT's performance and that of human programmers.

In summary, while ChatGPT represents a significant breakthrough in language modeling, its limitations in programming tasks suggest that there is still much room for improvement. Further research and development are needed to improve ChatGPT's performance on programming tasks, especially for more complex problems, and to bridge the gap between ChatGPT's performance and that of human programmers.

\subsection{Implications for AI development and applications}

The implications of theses results for AI development and applications in the programming field are significant. While ChatGPT and other language models have shown promise in natural language processing and generation, their limitations in complex programming tasks indicate that they may not be suitable for fully automated programming, at least not yet. However, they can still be useful for tasks such as generation of simple programs, code completion, code summarization, and documentation generation.

To fully leverage the potential of language models in programming, further research is needed to develop models that can understand and reason about code in the same way as human programmers. This will require a better understanding of the cognitive processes involved in programming and the ability to incorporate this knowledge into AI models. Additionally, more comprehensive and diverse datasets need to be developed that better capture the variety of programming tasks and languages used in real-world programming.

Overall, the limitations of ChatGPT in programming tasks highlight the need for continued research and development in AI and programming, and the importance of understanding the strengths and limitations of AI models in different contexts.
\section{Conclusion}

Numerous studies have demonstrated the impressive performance of ChatGPT, which often rivals or even surpasses human capabilities in various tasks and domains. However, this paper presented an alternative perspective by showing a situation where human performance excels over ChatGPT in typical tasks that suit it, specifically in relatively complex computer programming. To evaluate this claim quantitatively, we used the IEEExtreme Challenge competition as a benchmark, which offers a range of programming problems with varying levels of difficulty. We executed a diverse set of 102 challenges drawn from five IEEExtreme editions, using three major programming languages: Python, Java, and C++. We then compared ChatGPT's score to the average score achieved by human competitors. 

Our empirical analysis demonstrated that human programmers maintain a significant advantage over ChatGPT in certain aspects of problem-solving within the programming context. This paper offers critical insights into the potential areas of improvement for ChatGPT and other AI-based language models. Future research could investigate the factors that enable humans to outperform ChatGPT in programming tasks and explore ways to address the limitations of AI-based language models in this area, such as improving their understanding of programming languages and their ability to work with complex code structures. 



\bibliographystyle{IEEEtran}
\bibliography{references}

\end{document}